\documentclass[prd,showpacs,superscriptaddress,preprintnumbers,amsmath,nofootinbib]{revtex4}
\usepackage[dvips,final]{graphicx}
\usepackage{amssymb,amsmath,epsfig,bm,pifont,amsfonts}
\usepackage[utf8]{inputenc}
\newcommand{\be}{\begin{equation}}\newcommand{\ee}{\end{equation}}%
\newcommand{\bd}{\begin{displaymath}}\newcommand{\ed}{\end{displaymath}}
\newcommand{\bit}{\begin{itemize}}                                        
 \newcommand{\eit}{\end{itemize}}                                         
\newcommand{\ben}{\begin{enumerate}}                                      
 \newcommand{\een}{\end{enumerate}}                                       
\newcommand{\baa}{\begin{array}{lll}}                                     
 \newcommand{\eaa}{\end{array}}                                           
\newcommand{\ba}{\begin{eqnarray}}                                        
 \newcommand{\ea}{\end{eqnarray}}                                         
\newcommand{\la}{\label}                                                  
\newcommand{\Ds}{\displaystyle}                                           
\newcommand{\gev}[1]{\relax\ifmmode{\text{GeV}^{#1}}                      
                     \else{GeV$^{#1}${ }}\fi}                             
\def\MSbar{\relax\ifmmode\overline                                        
            {\rm MS}\else{$\overline{\rm MS}${ }}\fi}                     
\def\as{\relax\ifmmode \alpha_s\else{$ \alpha_s${ }}\fi}                  
\def\abar{\relax\ifmmode{\bar{a}}\else{$\bar{a}${ }}\fi}                  
  \def\ie{\hbox{\it i.e.}{ }}                 
    
\newcommand{\dAA}{d_A^{abcd}d_A^{abcd}}
\newcommand{\dRR}{d_F^{abcd}d_F^{abcd}}
\newcommand{\dRA}{d_F^{abcd}d_A^{abcd}}                                           
\newcommand{\dRRNA}{\frac{d_F^{abcd}d_F^{abcd}}{N_A}}
\newcommand{\dRANA}{\frac{d_F^{abcd}d_A^{abcd}}{N_A}}
\newcommand{\dAANA}{\frac{d_A^{abcd}d_A^{abcd}}{N_A}}

\usepackage{color}
\definecolor{mBlue}{rgb}{0,0,1}
 
\definecolor{mRed}{rgb}{1,0,0}
 
\definecolor{mGreen}{rgb}{0.0,1.0,1.0}
 
    \definecolor{DarkGreen}{rgb}{0.04,0.5,0.1}

\begin{document}
\thispagestyle{empty}
 \date{\today}

\title{
Optimized determination of the polarized Bjorken sum rule in pQCD}


\author{D. Kotlorz}
\email{dorota@theor.jinr.ru}
\affiliation{Department of Physics,
                Opole University of Technology,\\
                45-758 Opole, Proszkowska 76, Poland}
\affiliation{Bogoliubov Laboratory of Theoretical Physics, JINR,\\
                141980 Dubna, Russia}

\author{S. V. Mikhailov}
\email{mikhs@theor.jinr.ru}
\affiliation{Bogoliubov Laboratory of Theoretical Physics, JINR,\\
                141980 Dubna, Russia}
\begin{abstract}
We present the method of numerical optimization for the perturbative series
using the renormalization group  in quantum chromodynamics.
We apply our approach to the perturbation series in $\alpha_s$ for the
coefficient function $C_\text{Bjp}(\alpha_s)$ of the Bjorken sum rule
for the polarized deep inelastic lepton-hadron scattering.
We optimize the Bjorken sum rule value, $\Gamma_1^\text{p-n}$, at the COMPASS,
SLAC and JLab kinematics and compare the obtained results with the experimental measurements
and also with the truncated Bjorken sum rule  predictions.
\end{abstract}
\pacs{11.55.Hx, 12.38.-t, 12.38.Bx}
\pdfinfo{%
  /Title    ()
  /Author   ()
  /Creator  ()
  /Producer ()
  /Subject  ()
  /Keywords ()
}

\maketitle
\section{Introduction}
\label{sec:intro}
One of the foundations of quantum chromodynamics (QCD) is the renormalization
group (RG) equation. The latter defines the running of the coupling $\alpha_s(Q^2)$
that characterizes the strong interaction 
of quarks and gluons.
Understanding the behavior of $\alpha_s$ with the scale of the virtual momenta
$Q^2$ allows us to describe hadronic interactions at both short and long
distances. The long-distance domain at low $Q^2$ is characteristic of quark
confinement and processes of hadronization, and the short-distance domain, at
high $Q^2$, involves perturbative methods of QCD (pQCD). The calculation
of the 1-loop $\beta$-function in pQCD over 40 years ago enabled
the discovery of asymptotic freedom. Since then, tremendous progress has
been made in perturbative calculations in QCD. Particularly, the Bjorken
sum rule (BSR) for the polarized deep inelastic lepton-hadron scattering
(DIS) \cite{Bjorken:1966jh, Bjorken:1969mm},
providing fundamental spin predictions of the nucleon, has been studied in
detail, both theoretically and experimentally. The radiative corrections to
BSR in the strong coupling constant $\alpha_s$ of order
$O(\alpha_s^n)$, $n=1,..,4$ were obtained in
\cite{Kodaira:1978sh,Gorishnii:1985xm,Larin:1991tj,Baikov:2010je},
respectively.
In order to optimize the perturbative series in $\alpha_s$ of a physical
observable, various methods can be used. One of them is the
Brodsky-Lepage-Mackenzie (BLM) approach \cite{Brodsky:1982gc},
later on developed in two different approaches:
the sequential BLM (seBLM), e.g., \cite{Mikhailov:2004iq,Kataev:2014jba},
and the principle of maximal conformality (PMC),
e.g., \cite{Deur:2017cvd} and references therein.

In this paper, we present another method of optimization applied to
the coefficient function $ C_\text{Bjp} (\alpha_s)$ for BSR predictions.
In the next section, we discuss and fix the criteria for the optimized analysis
of the $C_\text{Bjp}(\alpha_s)$ within the perturbation QCD and RG.
Following these criteria, in Sec.~\ref{sec:sec5}, we find numerically
the admissible domains for the corresponding new normalization scales
$\mu^2$ ($\mu^2 \neq Q^2$).
In Sec.~\ref{sec:sec6}, the results of optimization are presented and discussed.
We optimize the perturbation expansion for the BSR at the COMPASS, SLAC and Jefferson Lab
(JLab) kinematics and compare the obtained results with the experimental
measurements and also with our predictions based on the truncated Bjorken
sum rule (tBSR) approach \cite{Kotlorz:2017wpu,D.Kotlorz:2019oyu}.

\section{Renormalization group analysis of QCD PT series for Bjorken sum rule}
\label{sec:sec4}
For decades, perturbative QCD has been a powerful tool in understanding the
hadron structure. In the analysis, the hadronic observables, like DIS sum rules, are expanded into
the power series in the strong coupling $\alpha_s$, providing a robust test
of the perturbative theory (PT). The Bjorken sum rule \cite{Bjorken:1966jh, Bjorken:1969mm},
$\Gamma_1^{p-n}(Q^2)$, which is a fundamental sum rule for polarized DIS
at hard momentum transfer $q: -q^2=Q^2$ and a rigorous
pQCD prediction, is essential for describing the nucleon spin structure.
In the limit $Q^2\to\infty$, the BSR relates the difference
between the first moments of the proton, $g_1^p$, and the neutron, $g_1^p$,
spin structure functions to the nucleon axial charge, $g_A$,
$\Gamma_1^{p-n}=|g_A|/6$. Since in real experiments $Q^2$ cannot reach infinity,
the QCD analysis of the BSR involves both the perturbative leading-twist and
the nonperturbative higher-twist (HT) corrections. Thus, away from the
large $Q^2$ limit, the $Q^2$ dependence of the polarized BSR is given by
\be
\Gamma_1^\text{th}(Q^2) = \Gamma_1^p(Q^2)-\Gamma_1^n(Q^2) =
\bigg{|}\frac{g_A}{6}\bigg{|}\, C_\text{Bjp}(a_s) +
\sum_{i=2}^{\infty}\frac{\mu_{2i}^{p-n}}{Q^{2i-2}}\, , \label{eq.4.1a}
\ee
where $C_\text{Bjp}(a_s)$ is the leading-twist nonsinglet coefficient function (c.f.)
including radiative corrections obtained within the \MSbar scheme and is
known to 4 loops
\cite{Kodaira:1978sh,Gorishnii:1985xm,Larin:1991tj,Baikov:2010je}.
The HT contribution is a series in power of $1/Q^2$, where $\mu_{2i}^{p-n}$
are the effective scales of the power corrections whose effects become essential in the small
and moderate $Q^2$ region.
Below, we will investigate the QCD radiative corrections to $C_\text{Bjp}(a_s)$ based on the renormalization group transform.

 \subsection{The problem of PT optimization for  coefficient function $C_\text{Bjp}(a_s)$}
  \label{sec:sec4a}
The perturbation expansion for the c.f. $C_\text{Bjp}(a_s)$ reads

\begin{subequations}
 \label{eq:4.1}
  \ba
\!\!\!\!\! C_\text{Bjp}\left(\frac{Q^2}{\mu^2},a_s(\mu^2)\right)&=&
1+c_1\left(a_s(\mu^2)+c_2~a_s^2(\mu^2)+ c_3~a_s^3(\mu^2)+c_4~a_s^4(\mu^2)+\ldots \right)\label{eq:4.1b}\,,
\ea
where the coefficients $c_i=c_i(Q^2/\mu^2)$ are calculated in the \MSbar scheme and
 are normalized
by the first coefficient $c_1=-3C_\text{F}=-4$; the running QCD coupling
$a_s$ is $ a_s(\mu^2) =\bm{\alpha_s}(\mu^2)/(4\pi)$.
For the default  condition  $\mu^2=Q^2$, the coefficients $c_i\equiv c_i(1)$ are the numbers
presented in Appendix~\ref{App:A} in different forms,
\ba
\!\!\!\!\!C(a_s(\mu^2))\equiv C_\text{Bjp}(1,a_s(\mu^2)) &=&
1-4\left(a_s(\mu^2)+ 13~a_s^2(\mu^2) + 221.6~a_s^3(\mu^2) + 6553.7~a_s^4(\mu^2)+\ldots \right)\,.\label{eq:4.1c}
\ea
  \end{subequations}
The numerical estimates in Eq.~(\ref{eq:4.1c}) are taken at the number of active quarks
$n_f=4$; see Eq.~(\ref{eq:C1-4numer}).
For the character reference scale of  BSR measurements
near the $\tau$-lepton mass, \mbox{$\mu^2=m_\tau^2 \approx 3.16$ GeV$^2$},
$a_s(m_\tau^2) \approx$ $0.332/(4\pi)\approx 0.0264$
[here $\alpha_s(m_\tau^2)=0.332 \pm 0.005$(exp) $\pm 0.015$(theor)
\cite{Baikov:2008jh}$\,$], 
one obtains for the series in Eq.~(\ref{eq:4.1c}) the estimate
\begin{subequations}
 \label{eq:3}
 \ba \label{eq:4.1d}
\!\!\!\!\!\!\!\!\!\!C(a_s(m_\tau^2))\equiv C_\text{Bjp}\left(1,a_s(m_\tau^2)\right)&=&1-
 4\left(0.0264+ 0.0090 + 0.0041+ 0.0032 +\ldots \right) \\
\!\!\!\!\!\!\!\!\!\! \phantom{C_\text{Bjp}(a_s)\equiv
C_\text{Bjp}\left(1,a_s(m_\tau^2)\right)}&=&1-4(~~0.0428~~)\,. \label{eq:4.1e}
 \ea
 \end{subequations}
One can see that the radiative corrections are significant, being about $-17$\% of the Born term in
Eq.~(\ref{eq:3}), and the convergence of the series is not very good.
Below, we perform \textit{an optimization} of the partial sum in Eq.~(\ref{eq:4.1b})
by choosing an appropriate new normalization scale $\mu \to \mu'$ and following the renormalization group  transform.
The value of the partial sum for the series in Eq.~(\ref{eq:4.1}) as well as the
values of its separate terms start to change
at the variation of the renormalization scale $\mu^2$ around the default  scale $Q^2$.
This is the inevitable effect of the series truncation which we will use for optimization.
Our goal is to make smaller the total value of radiative corrections in Eq.~(\ref{eq:4.1b}) keeping simultaneously
some natural hierarchy of the coefficients $c_i$ for appropriate convergence,
using for this purpose the variation of a scale $\mu$.
The corresponding approach goes back to the generalization of the BLM \cite{Brodsky:1982gc}
method, which was suggested in \cite{Mikhailov:2004iq,Kataev:2014jba} for the RG invariant quantities.
The approach is based on the $\{\beta \}$-expansion for the PT coefficients \cite{Mikhailov:2004iq}
\footnote{Another approach to construction of the $\{\beta \}$-expansion was suggested in \cite{Cvetic:2016rot},
but their results are numerically close.}; here they are presented for $c_i$ in Appendix \ref{App:A},
which allows us to derive the intrinsic structure of $c_i$ in connection with charge renormalization
in great detail.
An alternative approach to the PT optimization, named PMC (see \cite{Ma:2015dxa}), is elaborated and
applied to BSR in \cite{Deur:2017cvd}; we will discuss its results in Sec.~\ref{sec:sec6}.
However, it is not necessary to know all the details of the series structure to solve a practical optimization
of this series.
Here, we will avoid the details of $\{\beta \}$-expansion (different for different approaches)
and will not discuss them, but, instead, we propose
a direct numerical method to deal with the partial sums of the series.

In other words, following the RG transform, we will reorganize four successive orders of radiative corrections
in the parentheses in the rhs of Eq.~(\ref{eq:4.1d}) to make their sum minimal.
In the next subsection, we will remind the reader the required elements of the formalism for
transformation of the expansion coefficients for any RG invariant (RGI) quantity.

\subsection{Parametrization of the RG transformation}
\label{sec:sec4b}

We consider the transformation of the coefficients $c_i$ of the RGI quantity $C_\text{Bjp}(a_s)$ under the change
of the normalization scale $\mu \to \mu'$.
Let $a_s=\bar{a}_s(t)$ and $a'_s=\bar{a}_s(t')$ be the solutions of the RG equation for the QCD charge with
logarithmic argument $t=\ln(\mu^2/\Lambda_{qcd}^2)$ at the same integration constant $\Lambda_{qcd}^2$.
Reexpanding the running coupling $\bar{a}_s(t)= a_s(\Delta, a_s')$
in terms of  $\Delta=t-t'=\ln\left(\mu^2/\mu'^2\right)$ and the coupling $a_s'$, we obtain
\ba
\label{eq:RGrearrange}
a_s = a_s(\Delta, a_s')&=&
\exp\left[-\Delta \beta(\bar{a}_s)\partial_{\bar{a}_s}\right]\bar{a}_s\Bigr|_{\bar{a}_s=a_s'}=
a'_s - \beta(a'_s)\frac{\Delta}{1!} +
\beta(a'_s)\partial_{a'_s}\beta(a'_s)\frac{\Delta^2}{2!}
+
\ldots \,.
\ea
This is the way to write the  RG solution for $\bar{a}(t)$
through the operator
$\exp\left(-\Delta\, \beta(a)\partial_{a}\right)[\ldots]\mid_{a=a'}$
(see \cite{Mikhailov:2004iq,Kataev:2014jba} and references therein).
The shift $\Delta$ of the logarithmic scale in Eq.~(\ref{eq:RGrearrange})
can be expanded in its turn in the perturbation series in powers of the rescaled charge
$a_s'\beta_0$ \cite{Mikhailov:2004iq}:
\ba \la{Delta}
~t' &\equiv & t-\Delta, \nonumber \\
&&~~~~~ \Delta\equiv\Delta(a'_s)=\Delta_{0} + a'_s\beta_0~ \Delta_{1}
+ (a'_s\beta_0)^2~ \Delta_{2} + \ldots,
\ea
 where the argument of the new coupling $a'_s$ depends on $\Delta$, \ie,
 $a'_s \equiv \bar{a}_s(t')= a_s\left(t-\Delta(a'_s)\right)$.
Reexpansion  $a_s$ in terms of $a'_s$ and $\Delta_i$ leads to rearrangement of the perturbation series for
c.f. $C_\text{Bjp}(a_s)= \sum_i a^i_s c_i \to \sum_i (a'_s)^i c'_i$.
The new primed coefficients $c'_i$ there can be expressed as
$c'_i= B_{ij}c_j$, where $B_{ij}$ is a triangular matrix presented in Table~\ref{Tab:r_n.d_k}.
In this notation, $C_\text{Bjp}$ from Eq.~(\ref{eq:4.1c}) transforms to
\begin{equation}
C_\text{Bjp}(a_s) = \sum_{i\geqslant 0} a^i_s c_i \to \sum_{i\geqslant 0} (a'_s)^i c'_i = 1+\sum_{i,j\geqslant 1} (a'_s)^i B_{ij} c_j \,,
\label{eq:C-pt-final}
\end{equation}
when the normalization scale $\mu$ is transformed $\mu \to \mu' $.
The elements $B_{ij}$ appear as a composition of transforms in Eq.~(\ref{eq:RGrearrange})
taken together with the expansion in Eq.~(\ref{Delta}).

\begin{table}[h]
\caption{The first few elements of the matrix $B_{ij}$. New PT coefficients $c'_i=B_{ij}c_j$.
\label{Tab:r_n.d_k}}
\begin{tabular}{|p{40mm}|p{30mm}|p{30mm}|p{30mm}|}\hline
                                   &                              &                             & \\
                      \centerline{$\bm{1}$}
                                   &  \centerline{$0$}            & \centerline{$0$}            &\centerline{$0$}
\\ \hline
                                   &                              &                             & \\
          \centerline{$-\beta_0 \Delta_{0}$}
                                   & \centerline{$\bm{1}$}
                                     & \centerline{$0$}           & \centerline{$0$}
                                     \\ \hline
            \centerline{$\Ds-\beta_0^2\vphantom{^{\big|}}\Delta_{1}\vphantom{^{\big|}_{\big|}}$}
                                  $\Ds-\beta_1\vphantom{^{\big|}}\Delta_{0}\vphantom{^{\big|}_{\big|}}$
                                  $\Ds+\beta_0^2\vphantom{^{\big|}}\Delta_{0}^2\vphantom{^{\big|}_{\big|}}$
                                   &\centerline{$-2\beta_0 \Delta_{0}\vphantom{^{\big|}}$}
                                   & \centerline{$\bm{1}$}
                                                                                &  \centerline{$0$}
\\ \hline
                     \centerline{$\Ds-\beta_0^3\vphantom{^{\big|}}\Delta_{2}\vphantom{^{\big|}_{\big|}}$}
                     $-\beta_2\Delta_{0}-\beta_0\beta_1\Delta_{1}\vphantom{^{\big|}}$
                     $\Ds+\frac{3}2\beta_0\beta_1\Delta_{0}+2\beta_0^3\Delta_{0}\Delta_{1}\vphantom{_{\big|}}$
                                   &  \centerline{$\Ds-2 \beta_0^2\vphantom{^{\big|}}\Delta_{1}\vphantom{^{\big|}_{\big|}}$}
                                  $\Ds-2 \beta_1\vphantom{^{\big|}}\Delta_{0}\vphantom{^{\big|}_{\big|}}$
                                  $\Ds+3 \beta_0^2\vphantom{^{\big|}}\Delta_{0}^2\vphantom{^{\big|}_{\big|}}$
                                                & \centerline{$-3\beta_0 \Delta_{0}\vphantom{^{\big|}}$}
                                                             &\centerline{$\bm{1}$}
\\ \hline
\end{tabular}
\end{table}
Below, in the square brackets we write explicitly the elements of the triangle matrix $B$:
\begin{subequations}
\label{eq:rearrange}
  \ba
\!\!\!\!a_s^1\cdot c_1 \to&a_s'^1\cdot [c'_1=& 1]; \nonumber  \\
\!\!\!\!a_s^2\cdot c_2 \to&a_s'^2\cdot \big[c'_2(\Delta_{0})=&c_2 -1\cdot \beta_0 \Delta_{0}\big]; \la{1-stage-2} \\
\!\!\!\!a_s^3\cdot c_3 \to&a_s'^3\cdot \Big[c'_3(\Delta_{0},\Delta_{1})=&c_3 -c_2\cdot 2\beta_0\Delta_{0} -1\cdot \left(\beta_1\Delta_{0}- \beta_0^2\Delta_{0}^2+\beta_0^2\Delta_{1}\right)\bigg];
 \la{1-stage-3}\\
\!\!\!\!a_s^4\cdot c_4 \to&\!\!\!a_s'^4\cdot \Big[c'_4(\{\Delta_{i}\}_0^2)=&c_4 -
c_3\cdot 3\beta_0\Delta_{0} -c_2\cdot \left(2\beta_1\Delta_{0}-3\beta_0^2\Delta_{0}^2+2\beta_0^2\Delta_{1}\right) - \la{1-stage-4}\\
&&  -1\cdot\left(\beta_2\Delta_{0}+\beta_0\beta_1\Delta_{1}
-\frac{3}2\beta_0\beta_1\Delta_{0}^2- 2\beta_0^3\Delta_{0}\Delta_{1}+\beta_0^3\Delta_{2}\right)
\bigg]; \nonumber \\
\ldots&&\ldots \nonumber \\
\!\!\!\!a_s^n\cdot c_n \to&a_s'^n\cdot \Big[c'_n(\{\Delta_{i}\}_0^{n-2})=&c_n -
c_{n-1}\cdot (n-1)\beta_0\Delta_{0} -\ldots \Big]\,.
\ea
 \end{subequations}
New coefficients $c'_n$ in Eq.~(\ref{eq:rearrange}) depend on the fitted parameters $\Delta_{i}$.
The different approaches based on different  $\{\beta \}$-expansion for $c_i$ tell us how
to deal with $\Delta_{i}$ to fix these $c'_i$.
At this point it is instructive to recall the standard BLM~\cite{Brodsky:1982gc} procedure
which deals with $O(a_s^2)$ order and is based on the decomposition  $c_2=\beta_0 \cdot c_2[1] + c_2[0]$.
BLM fixes the scale $\Delta_{0}$ in Eq.~(\ref{1-stage-2}) by the requirement
$\Delta_{0}=c_2[1]$, and thereby $c_2 \to c'_2= \beta_0\cdot 0+ c_2[0]$.
This condition transfers 1-loop renormalization of charge, accumulated in the term $a_s^2 \beta_0 c_2[1]$,
into the new renormalization scale $\mu'^2=\exp(-c_2[1])\mu^2$ of the coupling
$a'_s$, where $\ln(\mu'^2/\mu^2)=t'-t=-\Delta_{0}$.
At the same time, the coefficient $c'_2$ is reduced to $c'_2=c_2[0]$, the ``conformal part'' of $c_2$.
For further generalization of the BLM approach one needs to know about the tracks of charge renormalization in
every higher order coefficient $c_i$, which are described by the so-called
$\{\beta \}$-expansion \cite{Mikhailov:2004iq,Kataev:2014jba} and which are presented in Appendix \ref{App:A}.

Another approach is to fit the parameters $\{\Delta_{0},\Delta_{1},\Delta_{2},\ldots \} \equiv \{\bm{\Delta}\}$ numerically
following some criteria of the PT series optimization and ignoring the intrinsic structure of $c_n$.
One can manage both the values of the PT coefficients $c'_i(\{\bm{\Delta} \})$ and the value of
the new coupling $a'_s=\bar{a}_s(t'=t-\Delta)$, and thereupon improve the convergence of expansion in the set of
Eqs.(\ref{eq:rearrange}).
By the same procedure, by means of Eqs.~(\ref{Delta}) and
(\ref{eq:rearrange}), we find a way to improve perturbation expansion.
The price we pay to achieve this improvement is that we have to \textit{control simultaneously} both
the expansion for $\Delta$ in Eq.~(\ref{Delta}) and for the coefficients $c'_i$ in Eq.~(\ref{eq:rearrange}).
In the paper, we develop just this approach.
In the next section, we will formulate the foregoing conditions and discuss the admissible domains of
$\{\Delta_{0},\Delta_{1},\Delta_{2}\ldots \}$ following from them.

\section{The admissible  domains of $\{\bm{\Delta} \}$ parameters}
\label{sec:sec5}
Let us apply the general scheme of optimization described in the previous section to the relevant quantity
$C_\text{Bjp}$ starting from the appropriate conditions for the truncated PT series in Eq.~(\ref{eq:C-pt-final}).
At first sight, it might seem that one can choose any value for the new scale $\mu'$ and, therefore,
the parameters $\{\Delta_{0},\Delta_{1},\Delta_{2},\ldots \}$ in Eq.~(\ref{Delta})
might look unconstrained, but that is not true.
In order to satisfy the reliability requirements for the PT expansion, we demand natural inequalities (i -- iii)
for its successive terms: \\
(i) The terms of PT expansion of $\Delta(t)$ in Eq.~(\ref{Delta}) should be
\be \label{eq:Delta-cond}
|\Delta_{0}|\geqslant |A'\Delta_{1}|\geqslant |A'^2 \Delta_{2}|\,,\;\;\;\;\; \text{where}\;\; A'\equiv\beta_0 a'_s\;,
\ee
which means that the next term of this PT expansion cannot be larger than the previous one.
These inequalities suppose nonlinear conditions for $\Delta_i$, which become more restrictive
for the case $\Delta_0>0$ by virtue of asymptotic freedom, $A' \simeq 1/t'$, where $t'$
is defined in Eq.~(\ref{Delta}).\\
(ii) For PT expansion in Eq.~(\ref{eq:C-pt-final}) we impose conditions with respect to $c'_i$,
which are similar to Eq.~(\ref{eq:Delta-cond}):
\be \label{eq:4.1-cond}
1 \geqslant \bigg{|}A'\frac{c'_2}{\beta_0}\bigg{|} \geqslant \bigg{|}A'^2\frac{ c'_3}{\beta_0^2}\bigg{|}
\geqslant \bigg{|}A'^3\frac{ c'_4}{\beta_0^3}\bigg{|}\,.
\ee
We assume Eqs.~(\ref{eq:Delta-cond}) and (\ref{eq:4.1-cond}) to be \textit{necessary conditions},
admitting at the same time that we can provide and substantiate more restricted ones.
The new coefficients $c'_i$ are given by Eq.~(\ref{eq:rearrange}),
while the explicit forms of the initial coefficients $c_{i}$ are presented in Appendix \ref{App:A}.
The running $\bar{a}_s$ has an asymptotic expansion, Eq.~(\ref{eq:beta.new.4L}) of Appendix \ref{App:B}, or
can be taken from the numerical solution of Eq.~(\ref{eq:beta3.new}).\\
(iii) To fix the PT domain of applicability, we put for the logarithmic variable $t'=t-\Delta(t')$
the appropriate lower bound at $\mu_0^2 \simeq 1$ GeV$^2$ that corresponds to
$t_{\mu_0}=\ln\left(\mu_0^2/\Lambda_{qcd}^2\right)\simeq 2.3$ at $\Lambda_{qcd}=\Lambda^{(n_f=4)}_{(4)}=0.318$\,GeV:
\ba \label{eq:PTbound}
t, t'\geqslant t_{\mu_0}&&  \Rightarrow t-2.3 \geqslant \Delta(t')=\Delta_{0} + A' \Delta_{1} + A'^2 \Delta_{2}\,.
\ea
Next, we will scan $t$ in the practically interesting interval $2.3 < t \leqslant 8$
($1 < \mu^2 \leqslant 301$~GeV$^2$) and we will localize at every $t$ the region of
the parameters $\{\Delta_{0},\Delta_{1},\Delta_{2}\}$, where the constraint
conditions, Eqs.~(\ref{eq:Delta-cond}), (\ref{eq:4.1-cond}) and (\ref{eq:PTbound}), are fulfilled simultaneously.
These conditions form the admissible domain in the $\{\bm{\Delta}\}$-space at every value of $t$,
denoted as $\{\bm{\bar{\Delta}}\}$, where one can perform optimization.\\
\begin{figure}[hbt]
\centering
 \includegraphics[width=0.50\textwidth]{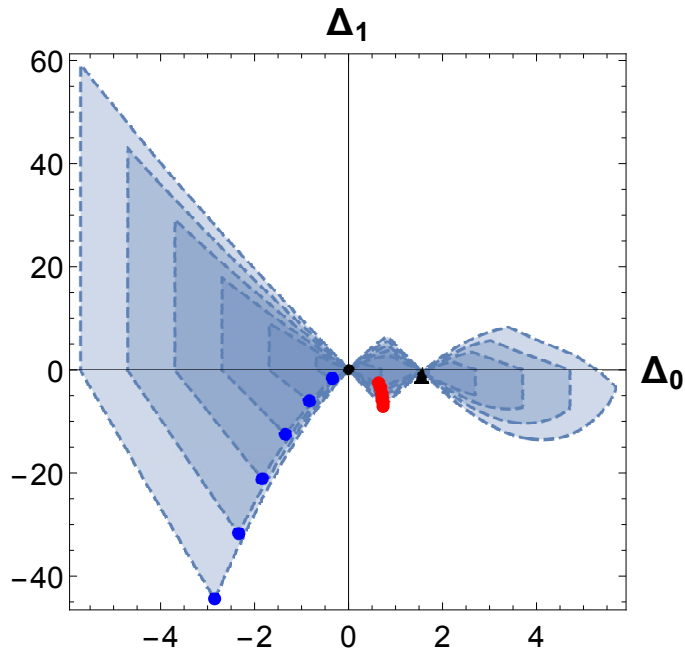}
\caption{
2D domains for admissible parameters $\{\bar{\Delta}_0, \bar{\Delta}_1\} $
at different $t$, distinguished by different degrees of gray: $t=3$ (dark), $\,\ldots\,$, $t=8$ (light).
The black triangle on the right half plane corresponds to the conditions $c'_2=c'_3=0$.
Blue points (on the left) and red points (on the right) are the bare
(global) and the constrained (local) minima of the radiative corrections, respectively.
\label{fig4}}
\end{figure}

\textbf{2D optimization}\\

\begin{figure}[ht]
\centering
 \includegraphics[width=0.4\textwidth]{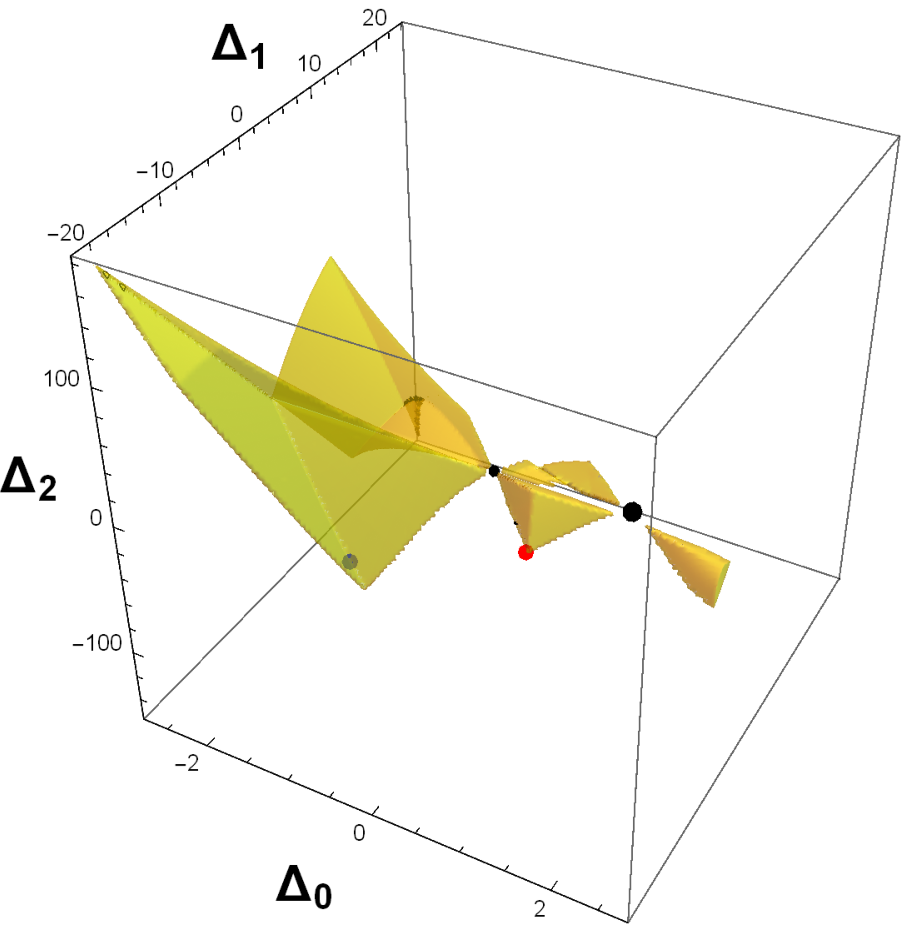}
 ~ \includegraphics[width=0.4\textwidth]{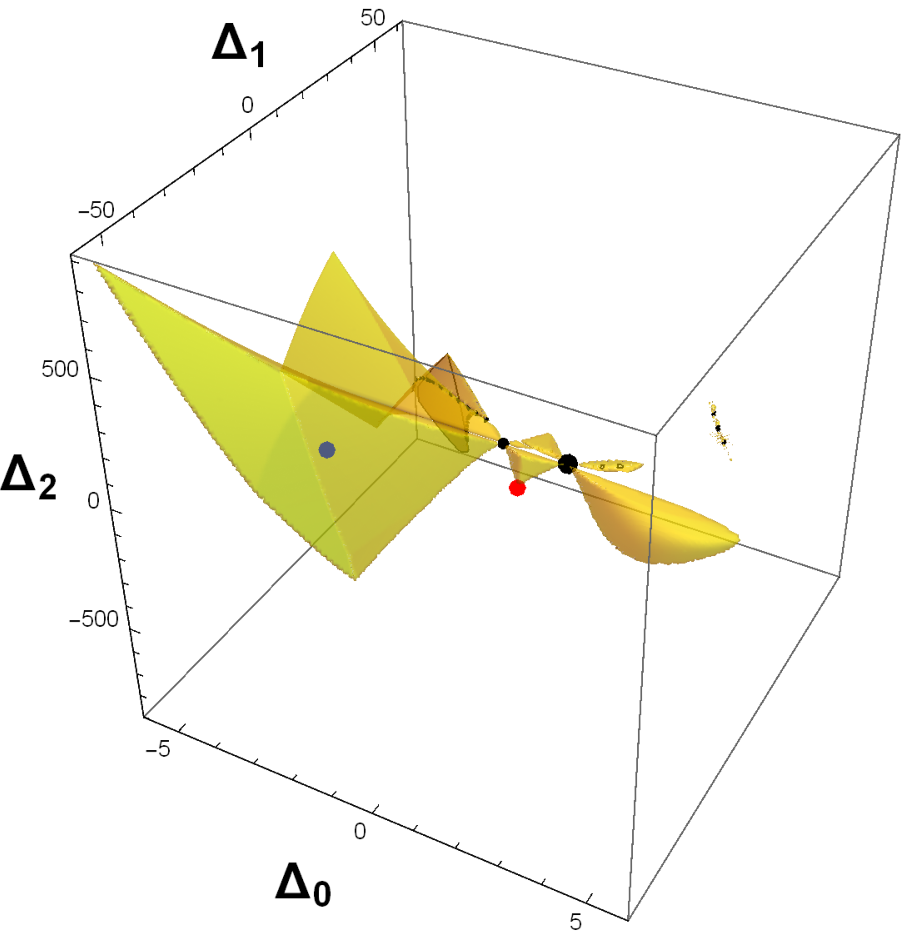}
\caption{
3D domains for admissible parameters $\{\bar{\Delta}_{0},\bar{\Delta}_{1},\bar{\Delta}_{2}\}$ at $t=5$ (left)
and $t=8$ (right).
The large black ball on the right half plane corresponding to the conditions $c'_2=c'_3=c'_4=0$
does not belong to the admissible region for both the cases.
Blue ($\Delta_0<0$) and red ($\Delta_0>0$)) points within the domains denote the bare
(global) and the constrained (local) minima of the radiative corrections, respectively, similarly as in Fig.~\ref{fig4}.
\label{fig5}}
\end{figure}

We consider first $C_\text{Bjp}$, Eq.~(\ref{eq:C-pt-final}), of order $O(a_s^3)$
for two-dimensional parametrization, $\{\Delta_{0},\Delta_{1}\}$. In this case,
the conditions, Eqs.~(\ref{eq:Delta-cond}), (\ref{eq:4.1-cond}) and (\ref{eq:PTbound}),
obtain the shortened form
\begin{subequations}\label{2dim}
 \ba
&|\Delta_{0}|\geqslant |A'\Delta_{1}|\, ,&  \label{2dima} \\
&1 \geqslant \bigg{|} \Ds A'\frac{c'_2}{\beta_0}\bigg{|} \geqslant\bigg{|} \Ds A'^2\frac{c'_3}{\beta_0^2}\bigg{|}\, ,& \label{2dimb} \\
&t \geqslant t_{\mu_0} + \Delta_{0} + A' \Delta_{1}\, .& \label{2dimc}
 \ea
\end{subequations}
The corresponding admissible domains calculated numerically for $t=3,\,4,\,\ldots\, ,\,8$ or, respectively,
for $\mu^2=2.0,\, 5.5,\, 15.0,\, 40.8,\,111,\,301$~GeV$^2$ at $\Lambda_{(4loop)}^{(n_f=4)} = 0.318$~GeV
are shown in Fig.~\ref{fig4}.
The constraints in Eqs.~(\ref{2dimb}) and (\ref{2dimc}) are much more restrictive for the
parameters in the right half plane for $\Delta_0>0$. Therefore, the corresponding domains are significantly
smaller than in the left half plane, where $\Delta_0<0$.
It is worth noting that the ``standard'' BLM value, $\{\Delta_0=2, \Delta_i=0 \}$, \cite{Brodsky:1982gc}
also belongs to the admissible domain.
Moreover, the larger $t$ is, the larger the corresponding admissible domain that is the manifestation of asymptotic freedom.
The point $(0,\,0)$ in the $(\Delta_0,\, \Delta_1)$-plane in Fig.~\ref{fig4} corresponds to the nonoptimized result
of $C_\text{Bjp}$, Eq.~(\ref{eq:4.1b}), while the points corresponding to the optimized one, Eq.~(\ref{eq:C-pt-final}),
are depicted for both $\Delta_0<0$ (left, blue circles) and $\Delta_0>0$ (right, red circles).
The black triangle, also lying in the admissible area, represents the conditions $c'_2=c'_3=0$
studied in \cite{Kataev:2014jba},
\ba \label{eq:c2=c3=0 }
c'_2=c'_3=0 \Rightarrow \Delta_{0}=c_2/\beta_0=1.56,~\Delta_{1}=
\left(c_3-c^2_2-c_2\beta_1/\beta_0 \right)/\beta^2_0 \approx -0.396\,.
\ea
The conditions in Eqs.(\ref{eq:c2=c3=0 }) correspond to the new norm scale
$\mu'^2 \!=\! \mu^2\! \exp\!\left[-\Delta(a'_s)\! =\! -1.56 +0.396 \beta_0 a_s(\mu'^2)\right]\!>\! 0.22 \mu^2$
see Fig.~1 (right) in \cite{Kataev:2014jba}.
If one imposes also the fourth term in the condition (ii), Eq.~(\ref{eq:4.1-cond}), for
the 2D parametrization, $\{\Delta_0, \Delta_1 \}$, the domains $\{\bm{\bar{\Delta}}\}$ in the right half plane
become slightly disintegrated and get a cut along the $\Delta_0$ direction.
This effect can be seen on the corresponding cross section at $\Delta_2=0$ of 3D admissible domains presented
in Fig.\ref{fig5}.\\

\textbf{3D optimization}\\

Next we consider the admissible 3D domains $\{\bar{\Delta}_0, \bar{\Delta}_1, \bar{\Delta}_2\}$
in the order of $O(a_s^4)$. Examples for $t=5~ (\mu^2 \approx 15.0~\text{GeV}^2 )$ and
$t=8~ (\mu^2 \approx 301.0~\text{GeV}^2 )$ are shown in Fig.~\ref{fig5} in the left and right panels, respectively.
We can see again that the larger $t$ is, the larger the corresponding admissible domain.
In comparison to the 2D case, 3D admissible domains become disintegrated for $\Delta_0>0$, having
a cut along the $\Delta_0$ direction.
Moreover, the large black balls in the 3D plots corresponding to the conditions $c'_2=c'_3=c'_4=0$, an analogue
of the 2D condition, Eq.~(\ref{eq:c2=c3=0 }), are not contained in the admissible regions.
The reason is obvious: this condition contradicts the inequality, Eq.~(\ref{eq:Delta-cond}),
for the perturbation expansion of ${\bm{\Delta}}$ for $t\leqslant 9$.

\section{The results of renormalization group optimization for BSR}
\label{sec:sec6}
In the previous sections, we have generally described the method of RG optimization.
Now we present the numerical results of this optimization for the Bjorken sum rule.
We consider optimization of c.f. $C_\text{Bjp}$ as a determination of the minimum of radiative corrections
by varying the parameters $\{\Delta_{0},\Delta_{1},\Delta_{2}\}$.\footnote{The problem of optimization of QCD radiative
corrections in the parameter space $\{\bm{\Delta} \}$ was formulated in \cite{Kataev:2014jba}.}
We find numerically the minimum of the function $|f_\text{Rad}|$ which
accumulates all of the known radiative corrections up to order $\alpha_s^4$,
\begin{subequations}
 \ba
 C_\text{Bjp}(t', a'_s)&=&1+ c_1 f_\text{Rad}(t;\{\bm{\Delta} \}) \equiv 1-\alpha^{g_1}_s/\pi\,, \label{eq5.1} \\
&&\phantom{1+ c_1}f_\text{Rad}(t;\{\bm{\Delta} \}) =
a_s' \left(1+a'_s c'_2+ (a'_s)^2 c'_3 + (a'_s)^3  c'_4 \right)\,. \label{eq5.2}
\ea
 \end{subequations}
The quantity $ |c_1 f_\text{Rad}(t;\{0 \})|= 4a_s \left(1+a_s c_2+ (a_s)^2 c_3 + (a_s)^3  c_4 \right)$
is the auxiliary ``effective $\alpha^{g_1}_s/\pi$'' charge proposed in \cite{Deur:2005cf,Deur:2017cvd}.
The arguments $\{\bm{\Delta}\}$ of $f_\text{Rad}(t;\{\bm{\Delta} \})$ are taken within the admissible domain
$\{\bm{\bar{\Delta}} \}$ at every $t$.
The approach of the PT optimization when the bare (global) minimum of $ |f_\text{Rad}(t;\{\bm{\Delta} \}|$ is restricted
by a set of inequality constraints, Eqs.~(\ref{eq:Delta-cond}), (\ref{eq:4.1-cond}) and (\ref{eq:PTbound}),
is universal and does not depend on the knowledge of details of the $\{\beta \}$-expansion for the quantity.
Indeed, in our numerical analysis, we will not need any information about the intrinsic structure of the
PT coefficients $c_i$.

\subsection{Numerical results of the renormalization group optimization of $C_\text{Bjp}$}
We compare different optimization results with the initial nonoptimized one, Eq.~(\ref{eq:4.1d}),
 \ba
 &&\{\Delta_{0}= \Delta_{1}=\ldots =0\} \nonumber \\
\!\!\!\!\!\!\!\!\!\!C_\text{Bjp}\left(1,a_s(m_\tau^2)\right)&=&1-
 4\left(0.0264+ 0.0090 + 0.0041+ 0.0032 +\ldots \right) \nonumber\\
\!\!\!\!\!\!\!\!\!\! \phantom{C_\text{Bjp}(a_s)\equiv
C_\text{Bjp}\left(1,a_s(m_\tau^2)\right)}&=&1-4(~~0.0428~~) = 1-\bm{0.171}=1-\alpha^{g_1}_s/\pi\,. \label{eq:5}
 \ea
Let us start with the 2D optimization in the $\{\Delta_{0}, \Delta_{1}\}$-space.
The optimal points (blue circles in Fig.~\ref{fig4}) are located for $\Delta_0<0$
on the boundary of admissible domains $\{\bar{\Delta}_{0}, \bar{\Delta}_{1}\}$.
For $t_\tau=\ln(m_\tau^2/\Lambda_{qcd}^2)\approx 3.44$ and $t'_\tau =  t_\tau - \Delta$ we have
\begin{subequations}
 \label{eqs:5.1}
 \ba
 &&\{\Delta_{0}=-0.571,\Delta_{1}=-3.35,\Delta_{2}=0 \} \label{eq:5.1} \\
\!\!\!\!\!\!\!\!\!\!C^{\Delta}_\text{opt,1}(a'_s)\equiv C_\text{Bjp}\left(t'_\tau,a'_s\right)&=&1-
 4\left(0.0205+ 0.0074 + 0.0054 + 0.0038 +\ldots \right) \label{eq:5.1a}\\
\!\!\!\!\!\!\!\!\!\! \phantom{C_\text{Bjp}(a_s)\equiv
C_\text{Bjp}\left(1,a_s(m_\tau^2)\right)}&=&1-4\,(~~0.0371~~)= 1 - \bm{0.149} =
1-\alpha^{g_1}_{s,\text{opt1}}/\pi\, . \label{eq:5.1b}
 \ea
 \end{subequations}
From the comparison between two cases: Eqs.~(\ref{eq:5}) and (\ref{eq:5.1b}),
we can see that the effective $\alpha^{g_1}_s/\pi$ changes from $\bm{ 0.171}$ to $\bm{ 0.149}$,
respectively, giving $\delta_\text{opt1} \equiv \alpha^{g_1}_s/\pi-\alpha^{g_1}_{s,\text{opt1}}/\pi \approx 0.023$.
The advantage of the optimized result looks substantial, while the values of the shift $\Delta$ in Eq.~(\ref{eq:5.1})
are moderate.

The next step is to find the minimum of $|f_\text{Rad}| $ at the additional condition, $\Delta_{0} > 0$.
This corresponds with the original BLM result, $\Delta_{0}^\text{BLM}=2$ [see Eq.(\ref{eq:C2})],
and also with similar results within the PMC \cite{Deur:2017cvd}.
The corresponding minima positions for $\Delta_0>0$, depicted as red circles in Fig.~\ref{fig4},
provide the following estimation:
\begin{subequations}
 \ba
\Delta_{0}>0 && \{\Delta_{0}=0.660,\Delta_{1}=-2.98, \Delta_{2}=0 \}  \label{eq:5.2}\\
\!\!\!\!\!\!\!\!\!\!C^{\Delta}_\text{opt,2}(a'_s)\equiv C_\text{Bjp}\left(t'_\tau,a'_s\right)&=&1-
 4\left(0.0266+ \underline{0.0053} + \underline{0.0053} + 0.0038 +\ldots \right)  \label{eq:5.2a} \\
\!\!\!\!\!\!\!\!\!\! \phantom{C_\text{Bjp}(a_s)\equiv
C_\text{Bjp}\left(1,a_s(m_\tau^2)\right)}&=&1-4\,(~~0.0410~~)=1 - 0.164\, . \label{eq:5.2b}
 \ea
 \end{subequations}
\noindent
This ``optimum'' result in Eq.~(\ref{eq:5.2b}) is not pronounced in comparison with Eq.~(\ref{eq:5.1b}).
At the same time, this solution appears on the boundary, where $(a')^2 c_2' \approx (a')^3 c_3'$;
see the underlined terms. This makes PT convergence worse and the final result less reliable.

For a similar 3D optimization within the admissible domains shown in Fig.~\ref{fig5} we obtain
\begin{subequations}
 \ba
 &&\{\Delta_{0}=-0.381,\Delta_{1}=-2.21,\Delta_{2}=-12.8 \} \label{eq:5.3} \\
\!\!\!\!\!\!\!\!\!\!C^{\Delta}_\text{opt,3}(a'_s)\equiv C_\text{Bjp}\left(t'_\tau,a'_s\right)&=&1-
 4\left(0.0207+ 0.0069 + \underline{0.0043} + \underline{0.0042} +\ldots \right) \label{eq:5.3a} \\
\!\!\!\!\!\!\!\!\!\! \phantom{C_\text{Bjp}(a_s)\equiv
C_\text{Bjp}\left(1,a_s(m_\tau^2)\right)}&=&1-4\,(~~0.0361~~)=1-0.144=1-\alpha^{g_1}_{s,\text{opt3}}/\pi
\label{eq:5.3b}
\ea
\end{subequations}
and
\begin{subequations}
 \ba
\Delta_{0}>0 && \{\Delta_{0}=0.573,\Delta_{1}=-2.72,\Delta_{2}=-5.69 \} \label{eq:5.4} \\
\!\!\!\!\!\!\!\!\!\!\!C^{\Delta}_\text{opt,4}(a'_s)\equiv C_\text{Bjp}\left(t'_\tau,a'_s\right)&=&1-
 4\left(0.0253+ 0.0053 + \underline{0.0045} + \underline{0.0044} +\ldots \right) \label{eq:5.4a}\\
\!\!\!\!\!\!\!\!\!\! \phantom{C_\text{Bjp}(a_s)\equiv
C_\text{Bjp}\left(1,a_s(m_\tau^2)\right)}&=&1-4\,(~~0.0396~~)=1-0.158\,. \label{eq:5.4b}
 \ea
  \end{subequations}
We see that the 3D analysis has no significant advantages over the corresponding 2D results,
Eqs.~(\ref{eq:5.1b}) and (\ref{eq:5.2b}). The underlined terms in Eqs.~(\ref{eq:5.3a}) and (\ref{eq:5.4a})
illustrate that the PT convergence is not good enough to make the results reliable.
Summarizing our tests among the considered cases, only the 2D result, Eq.~(\ref{eq:5.1b}),
is at a near optimum level providing satisfactory convergence of the PT series.

Let us comment on the disadvantage of the used numerical approach. Namely,
``blind analysis" based on the constraints, Eqs.~(\ref{eq:Delta-cond}), (\ref{eq:4.1-cond}), and (\ref{eq:PTbound}),
can lead to the unsatisfactory solution. Indeed, the minima of the radiative corrections in the cases,
Eqs.~(\ref{eq:5.2a}), (\ref{eq:5.3a}), and (\ref{eq:5.4a}), occur on the boundary of the
constraint, Eq.~(\ref{eq:4.1-cond})
(see the underlined terms), where the PT convergence deteriorates.
Another lesson from these numerical tests is that the scales of ``BLM/PMC," corresponding to the
additional condition $\Delta_0 > 0$, lead to the constrained (local) minimum which is not close to the
global minimum of radiative corrections.

In this connection, it is instructive to compare the results for $\alpha^{g_1}_{s\text{PMC}}(Q^2)/\pi$ obtained
in \cite{ Deur:2017cvd} within the PMC scale setting with our predictions.
The PMC is based on some version of $\beta$-expansion\footnote{Let us mention here that we do not
agree with a certain construction of $\{\beta \}$-expansion used in \cite{Deur:2017cvd}
for $c_{2,3,4}$; see our criticism in \cite{Kataev:2014jba,Kataev:2016aib}.}
and the assumption about sufficient convergence of the truncated PT series.
The conventional pQCD effective charge $\alpha^{g_1}_s/\pi > \alpha^{g_1}_{s \text{PMC}}/\pi$ at much better convergence
of the latter.
Let us briefly discuss similarities:
(i) The PMC scales qualitatively agree with the scales $\Delta$ obtained in
this paper for the local minimum $\Delta_0>0$ in Eqs.~(\ref{eq:5.2}) and (\ref{eq:5.4}).
(ii) The advantage of the PMC defined as
$\delta_\text{PMC}=(\alpha^{g_1}_s/\pi - \alpha^{g_1}_{s \text{PMC}}/\pi)\Bigr|_{Q^2=m^2_\tau} \approx 0.013$
(see Fig.~1 in \cite{Deur:2017cvd}) is numerically close to $\delta$ from Eq.~(\ref{eq:5.4b}) here.
In contrast, our 2D optimization for the global minimum, Eq.~(\ref{eqs:5.1}),
gives an almost twice higher advantage with appropriate convergence,
$\delta_\text{opt1} \approx 0.023$ vs $\delta_\text{PMC}$.
These differences originate from the fact that in our approach
the optimization is tightly related to the global minimum of the partial sum of radiative corrections,
as opposed to the PMC purpose of faster convergence.


\subsection{Optimized Bjorken sum rule vs different experimental results}
\label{sec:subsec6}
In this subsection, we compare the results for the BSR obtained here for the optimized PT series
for the leading twist contributions
with the experimental measurements of COMPASS \cite{Alekseev:2010hc, Adolph:2015saz, Adolph:2016myg}
(taking into account the results of truncated moments \cite{Kotlorz:2017wpu, Strozik-Kotlorz:2017gwn, D.Kotlorz:2019oyu}),
the E155 Collaboration at SLAC \cite{Anthony:2000fn}, and JLab EG1-DVCS \cite{Deur:2014vea}.
These optimized estimates are expectedly higher than the conventional ones.

\subsubsection{Optimized Bjorken sum rule for COMPASS measurements.}
Experimental verification of the DIS sum rules always encounters the difficulty that in any realistic experiment
one cannot reach arbitrarily small values of the Bjorken $x$, $x \geqslant x_0 \equiv Q^2_\text{min}/(2(Pq)_\text{max}>0)$,
where $P$ is hadronic momentum and $q$ the momentum transfer of the DIS.
The method of truncated Mellin moments (TMM) operating in the range $(x_0,1)$ can overcome this $x_0$ problem
\cite{Forte:1998nw, Kotlorz:2006dj, Kotlorz:2014fia}.
To obtain the optimized phenomenological result for BSR,
$\Gamma^\text{exp}_1 \simeq \Gamma_1(x_0)= \int^1_{x_0} g_1^{(ns)}(x;\mu^2)\, dx$,
we used the  tBSR  approach which incorporates experimental uncertainties on the spin function $g_1$
\cite{Strozik-Kotlorz:2017gwn, D.Kotlorz:2019oyu}.
The tBSR elaborated in \cite{Kotlorz:2017wpu} is based on
the TMM approach providing not only a natural framework of DIS analysis in the restricted kinematic region
of $ x \geqslant x_0$ but also allowing an effective study of the sum rules in a low $x$ limit.
Since the tBSR saturates in the low-$x$ limit much sooner than the ordinary BSR \cite{Kotlorz:2017wpu},
we assume a smaller systematic error and the total one of the level of $5\%$ at a conservative approach to this estimation.
Thus, we find for the COMPASS data
\be
\Gamma^\text{exp-opt}_{1(c-ss)}=0.191\pm 0.01\, ,
\label{res1_gBSR}
\ee
which is in good agreement with the most recent COMPASS result provided for $Q^2=3$~\rm{GeV}$^2$
\cite{Adolph:2016myg}:
\be
\Gamma^\text{exp}_{1(c-ss)}=0.192\pm 0.007_{\rm stat}\pm 0.015_{\,\rm syst}\, .
\label{res1_exp}
\ee

In the previous section, we discussed the optimized results for the QCD
radiative corrections at the world reference scale $m_\tau^2$.
Below, we provide similar results starting with $C_\text{Bjp}\left(1,a_s(Q^2)\right)$ at the COMPASS
reference scale $Q^2=3$~\rm{GeV}$^2$. Thus, we obtain
 \ba
\!\!\!\!\!\!\!\!\!\!C_\text{Bjp}\left(1,a_s(Q^2)\right)&=&1-
 4\left(0.0268+ 0.0093 + 0.0043+ 0.0034 +\ldots \right) \nonumber\\
\!\!\!\!\!\!\!\!\!\! \phantom{C_\text{Bjp}(a_s)\equiv
C_\text{Bjp}\left(1,a_s(3~\rm{GeV}^2)\right)}&=&1-4(~~0.0438~~) =
 1-0.175\,. \label{eq:compass}
 \ea

Then, using the already discussed and approved 2D $\{\Delta_0,\Delta_1\}$ optimization in Eq.~(\ref{eqs:5.1}),
we find the optimized value of $C_\text{Bjp}$,
 \begin{subequations}
 \ba
 &&\{\Delta_{0}=-0.545,\Delta_{1}=-3.13,\Delta_{2}=0 \} \label{eq:6.1} \\
\!\!\!\!\!\!\!\!\!\!C^{\Delta}_\text{opt}(a'_s)\equiv C_\text{Bjp}\left(t',a'_s\right)&=&1-
 4\left(0.0209+ 0.0077 + 0.0055 + 0.0039 +\ldots \right) \label{eq:6.1a}\\
\!\!\!\!\!\!\!\!\!\! \phantom{C_\text{Bjp}(a_s)\equiv
C_\text{Bjp}\left(1,a_s(m_\tau^2)\right)}&=&1-4(~~0.0380~~)=1-0.152\,, \label{eq:6.1b}
 \ea
 \end{subequations}
that is visibly larger than the nonoptimized result $1-0.175$ in Eq.~(\ref{eq:compass}).
These values lead to the following estimates for the leading twist-2 part of $\Gamma_1^\text{th}$
in Eq.~(\ref{eq.4.1a}):
\begin{subequations}
\ba
\Gamma_{1,tw2}^\text{th-non-opt}(Q^2) &=&
\bigg{|}\frac{g_A}{6}\bigg{|}_\text{C-SS}\, C_\text{Bjp}\left(1,a_s(Q^2)\right)\approx 1.29/6 \cdot 0.825
=0.177 \pm 0.003\, ,
\label{eq:6.2b} \\
\Gamma_{1,tw2}^\text{th-opt}(Q^2) &=&
\bigg{|}\frac{g_A}{6}\bigg{|}_\text{C-SS}\, C^{\Delta}_\text{opt}(a'_s)\approx 1.29/6 \cdot 0.848
=0.182 \pm 0.003\, ,
\label{eq:6.2a}
\ea
 \end{subequations}
where $|g_A|_\text{C-SS}=1.29\pm 0.05_\text{stat} \pm 0.1_\text{syst}$
is the specific estimate obtained together with $\Gamma^\text{exp}_{1(c-ss)}$ in
Eq.~(\ref{res1_exp}) in the recent COMPASS measurement \cite{Adolph:2016myg}.
The uncertainties of $\Gamma_{1,tw2}^\text{th}$ are determined by the uncertainty of $\alpha_s(m^2_\tau)$.
The result of the 3D optimization for the bare minimum, $\Gamma_{1,tw2}^\text{th}\approx 0.183$,
does not improve the estimate distinctly.
It is seen from the comparison of the nonoptimized result, Eq.~(\ref{eq:6.2b}), the optimized one, Eq.~(\ref{eq:6.2a});
and then the prediction of the tBSR approach, Eq.~(\ref{res1_gBSR}), with the experimental result, Eq.~(\ref{res1_exp}),
that the optimization reduces the differences between theoretical and experimental (exp, exp-opt)
estimations.

Let us now briefly discuss possible implementation of HT corrections in our
theoretical analysis.\\
At lower $Q^2$ the HT power corrections $\mu_4^{p-n}/Q^2$,
$\mu_6^{p-n}/Q^4$, etc., Eq.~(\ref{eq.4.1a}), are needed to describe the data.
The precise JLab data on $\Gamma_1^{p-n}(Q^2)$ at low $Q^2$ provided a good test
for both the perturbative leading-twist (LT) and nonperturbative HT contributions.
Several theoretical and experimental estimates of the scales $\mu_{2i}$ have been made showing
that impact of the HT corrections can be important for $Q^2\lesssim 5~\rm{GeV}^2$, see, e.g.
\cite{Balitsky:1989jb,Deur:2008ej,Pasechnik:2008th,Pasechnik:2009yc,Pasechnik:2010fg,Khandramai:2011zd,Deur:2014vea}.
After taking into account the first HT term $\mu_4^{p-n}$, which is negative,
the difference between our theoretical prediction and the COMPASS results will increase even more.
For the estimate $\mu_{4(kmtk)}^{p-n}/M^2= -0.047\pm 0.02$ from
\cite{Kotlorz:2017wpu}, we obtain
\be
\Gamma_{1}^\text{th-opt}(Q^2) = \Gamma_{1,tw2}^\text{th-opt}(Q^2) +
\frac{\mu_{4(kmtk)}^{p-n}}{Q^2} = 0.167 \pm 0.007\, ,
\label{eq:HT1}
\ee
while for the experimental estimate provided by JLab EG1-DVCS, \cite{Deur:2014vea},
$\mu_{4(JLab)}^{p-n}/M^2= -0.021\pm 0.016$, we have
\be
\Gamma_{1}^\text{th-opt}(Q^2) = \Gamma_{1,tw2}^\text{th-opt}(Q^2) +
\frac{\mu_{4(JLab)}^{p-n}}{Q^2} = 0.175 \pm 0.006\, ,
\label{eq:HT2}
\ee
where the uncertainty of $\Gamma_{1}^\text{th-opt}$ is the combined uncertainty from $\alpha_s$ and HT.
This latter result in Eq.~(\ref{eq:HT2}) is supported by the COMPASS value $\Gamma^\text{exp}_{1(c-ss)}$ in
Eq.~(\ref{res1_exp}) within combined statistical and systematic uncertainty and does not contradict even
more restricted $\Gamma^\text{exp-opt}_{1(c-ss)}$ in Eq.~(\ref{res1_gBSR}).\\


\subsubsection{Optimized Bjorken sum rule for SLAC E155 Collaboration measurements.}

The initial result for the coefficient function $C_\text{Bjp}$ at the reference scale $Q^2=5$~\rm{GeV}$^2$
of E155 measurements reads
 \ba
\!\!\!\!\!\!\!\!\!\!C_\text{Bjp}\left(1,a_s(Q^2)\right)&=&1-
 4\left(0.0236 + 0.0073 + 0.0029 + 0.0020  +\ldots \right) \nonumber\\
\!\!\!\!\!\!\!\!\!\! \phantom{C_\text{Bjp}(a_s)\equiv
C_\text{Bjp}\left(1,a_s(5~\rm{GeV}^2)\right)}&=&1-4(~~0.0358~~) =
 1-0.143=0.857\,. \label{eq:slac}
 \ea
Within the 2D $\{\Delta_0,\Delta_1\}$ optimization, Eq.~(\ref{eq:slac}), at $Q^2$ we find
 \begin{subequations}
 \ba
 &&\{\Delta_{0}=- 0.800 ,\Delta_{1}=-5.513,\Delta_{2}=0 \} \label{eq:22.a} \\
\!\!\!\!\!\!\!\!\!\!C^{\Delta}_\text{opt}(a'_s)\equiv C_\text{Bjp}\left(t',a'_s\right)&=&1-
 4\left( 0.0174 + 0.0060 + 0.0046 + 0.0029 + \ldots \right) \label{eq:22.b}\\
\!\!\!\!\!\!\!\!\!\! \phantom{C_\text{Bjp}(a_s)\equiv
C_\text{Bjp}\left(1,a_s(m_\tau^2)\right)}&=&1-4(~~ 0.0309 ~~)=1-0.124\, =\, 0.876\,, \label{eq:22.c}
 \ea
 \end{subequations}
while within the 3D optimization we obtain the result that is very close to the previous one
but with badly convergent PT series:
\begin{subequations}
 \ba
 &&\{\Delta_{0}=-0.548,\Delta_{1}=-3.58,\Delta_{2}=-24.3 \} \label{eq:23.a} \\
\!\!\!\!\!\!\!\!\!\!C^{\Delta}_\text{opt,5}(a'_s)\equiv C_\text{Bjp}\left(t'_\tau,a'_s\right)&=&1-
 4\left(0.0177+ 0.0055 + \underline{0.0035} + \underline{0.0035} +\ldots \right) \label{eq:23.b} \\
\!\!\!\!\!\!\!\!\!\! \phantom{C_\text{Bjp}(a_s)\equiv
C_\text{Bjp}\left(1,a_s(m_\tau^2)\right)}&=&1-4\,(~~0.0301~~)=1-0.120=0.880. \label{eq:23.c}
\ea
 \end{subequations}
Therefore, hereafter we take the 2D result, Eq.~(\ref{eq:22.c}), which gives the
leading-twist contribution
of BSR,
\be
\Gamma_{1,tw2}^\text{th-opt}(Q^2) =
\bigg{|}\frac{g_A}{6}\bigg{|}\,C^{\Delta}_\text{opt}(a'_s)= 1.27/6 \cdot  0.876=0.186 \pm 0.002\, ,
\label{eq:24}
\ee
where $|g_A|=1.2723 \pm 0.0023$ \cite{Tanabashi:2018}.
Including in the estimates of Eq.~(\ref{eq.4.1a}) also the HT contributions with the values
$\mu_{4(kmtk)}^{p-n}/M^2$ \cite{Kotlorz:2017wpu} and $\mu_{4(JLab)}^{p-n}/M^2$ \cite{Deur:2014vea},
we obtain
\begin{subequations}
\ba
\Gamma_{1}^\text{th-1}(Q^2) &=& \Gamma_{1,tw2}^\text{th-opt}(Q^2) +
\frac{\mu_{4(kmtk)}^{p-n}}{Q^2} = 0.177 \pm 0.004, \label{eq:25} \\
\Gamma_{1}^\text{th-2}(Q^2) &=& \Gamma_{1,tw2}^\text{th-opt}(Q^2) +
\frac{\mu_{4(JLab)}^{p-n}}{Q^2} = 0.182 \pm 0.003.\label{eq:27}
\ea
\end{subequations}

The optimized estimation in Eq.~(\ref{eq:25}) is in good agreement with the experimental result
\be
\Gamma_{1(SLAC)}^\text{exp}=0.176 \pm 0.003_\text{stat} \pm 0.007_\text{sys}\, ,
\label{eq:26}
\ee
while for the smaller in modulo HT correction in Eq.~(\ref{eq:27}) agreement to the data is reasonable.

\subsubsection{Optimized Bjorken sum rule for JLab EG1-DVCS measurements.}
\begin{figure}[h]
\centering
 \includegraphics[width=0.5\textwidth]{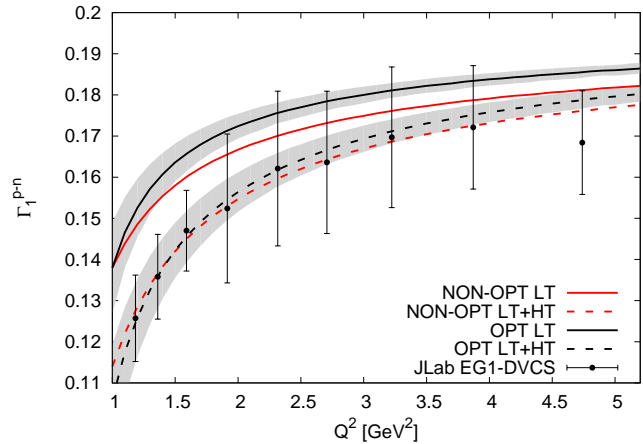}
\vspace{7mm} 
\caption{ \label{fig6} 
 Comparison of the optimized (black solid curve) and nonoptimized (red solid lower curve)
predictions on BSR with the experimental EG1-DVCS data. The impact of the
twist-4 correction is also shown (dashed).
For better visibility we show the error band only for the optimized plots. }
 \end{figure}

It is worthwhile to compare our analysis with the recent high precision determination of BSR at JLab \cite{Deur:2014vea}.
To this aim, we choose JLab EG1-DVCS data covering the range $1.0\leq Q^2\leq 4.8\,\rm{GeV^2}$
where the perturbative methods are justified. These experimental results are compared with
the optimized predictions of BSR in Fig.~\ref{fig6}.

We use twist-2 optimized values, Eqs.~(\ref{eq.4.1a}) and (\ref{eq5.2}), 
calculated for different experimental $Q^2>1\,\rm{GeV^2}$ together with the nonoptimized
ones, Eqs.~(\ref{eq.4.1a}) and (\ref{eq:4.1c}). 
For both cases we also estimated twist-4 correction from
the data and obtained $\mu_4^{p-n}/M^2=-0.034\pm 0.007$ for the optimized and
$-0.026\pm 0.007$ for the nonoptimized approach.
Figure~\ref{fig6} shows that the pure LT contribution to BSR lies
significantly above the experimental data for both kinds of theoretical results,
and the difference grows with decreasing $Q^2$,
motivating the necessity of HT corrections. It is found that the value $\mu_{4(opt)}^{p-n}/M^2=-0.034\pm 0.007$ estimated
from comparison of the optimized predictions to the data is compatible with
the experimental value provided by JLab EG1-DVCS and also with other theoretical
estimations. It is also seen that the optimized approach to the value LT$+$HT describes well the
$Q^2$ evolution of BSR even down to small $Q$ values.

\section{Conclusions}
\label{sec:concl}
We have discussed a possible improvement in theoretical determination of the polarized Bjorken sum rule $\Gamma_1^{p-n}$.
We performed minimization of the partial sums of the QCD perturbation series for the coefficient function
$C_\text{Bjp}(Q^2/\mu^2,\alpha_s(\mu^2))$ of the leading twist for a certain DIS process by means of an appropriate normalization scale $\mu^2$
resulting from the renormalization group. To this aim, we provided a set of general conditions for the optimized
analysis of $C_\text{Bjp}$ within the perturbation QCD and the renormalization group.
This frame is universal and applicable for the analysis of any renormalization group invariant quantities.
Based on these conditions, we found the admissible domains for the corresponding new normalization scales
$\mu^2$ for the cases of QCD corrections of the orders of $O(\alpha_s^3)$ and $O(\alpha_s^4)$.
For these domains we found numerically the minimum of the radiative corrections to $C_\text{Bjp}(\alpha_s)$
based on the 4-loop run of $\alpha_s(\mu^2)$. This leads to the optimum values of the theoretical predictions for BSR.
The optimized results for BSR in the order $O(\alpha_s^4)$ are systematically higher
than the standard ones and the difference varies between 0.006 at $Q^2=2\,\rm{GeV^2}$ and 0.003 at $Q^2=10\,\rm{GeV^2}$.
We compared these optimized results including also the essential twist-4 correction with the experimental measurements of
COMPASS \cite{Adolph:2016myg}, E155 \cite{Anthony:2000fn} and JLab EG1-DVCS \cite{Deur:2014vea}.
We obtained for COMPASS $\Gamma_{1}^\text{th-opt}(3\,\text{GeV}^2)= 0.175 \pm 0.006$ and for E155 SLAC
$\Gamma_{1}^\text{th-opt}(5\,\text{GeV}^2)= 0.177 \pm 0.004$.
Thus, for the precise E155 data we obtained good agreement with the experimental value
$\Gamma_{1(SLAC)}^\text{exp}=0.176 \pm 0.003_\text{stat} \pm 0.007_\text{sys}$ 
while for COMPASS data, which suffer from large statistical and experimental systematical uncertainties
compared to the SLAC E155 or JLab EG1-DVCS results, we obtained reasonable
agreement within the combined statistical and systematic uncertainty.
From comparison with the EG1-DVCS precise data for $Q^2>1\,\rm{GeV^2}$ we found that the optimized approach
to LT$+$HT describes well the $Q^2$ evolution of BSR even down to small $Q$ values.
Comparing the optimized predictions to the JLab data, we estimated the twist-4 correction
and obtained $\mu_{4(opt)}^{p-n}/M^2=-0.034\pm 0.007$ which is compatible with the experimental value 
$\mu_{4(JLab)}^{p-n}/M^2= -0.021\pm 0.016$ provided by
EG1-DVCS and also with other theoretical estimations.

\begin{acknowledgments}
We would like to thank  A.~L. Kataev and O.~V. Teryaev for the fruitful discussions and Y.~ Bedfer and E. M.~Kabus
for the communication of the data.
D. K. thanks A. Kotlorz for his help in numerical computations.
This work is supported by the Bogoliubov-Infeld Program.
D. K. acknowledges the support of the Collaboration Program JINR-Bulgaria 2019.
S. V. M. acknowledges the support of the BelRFFR-JINR, Grant No. F18D-002.
\end{acknowledgments}
\appendix
\section{$\{\beta \}$-expansion for $C_\text{Bjp}$}
 \renewcommand{\theequation}{\thesection.\arabic{equation}}
\label{App:A}   \setcounter{equation}{0}
\textbf{1.}
The $\{\beta\}$-expansion representation introduced in \cite{Mikhailov:2004iq}
prescribes the following form of decomposition of the perturbation coefficients $c_n$ for $C_\text{Bjp}$ in
Eq.~(\ref{eq:4.1}) or for any other RGI quantities:
\begin{subequations}
\label{eq:d_beta}
\begin{eqnarray}
\label{eq:c_1}
c_1&=&c_1[0]\, , \\
c_2&=&\! \beta_0\,c_2[1]
  + c_2[0]\, ,\label{eq:c_2}\\
  c_3
&=&\!
  \beta_0^2\,c_3[2]
  + \beta_1\,c_3[0,1]
  +       \beta_0 \,  c_3[1]
  + c_3[0]\, ,\label{eq:c_3} \\
  c_4
   &=&\! \beta_0^3\, c_4[3]
     + \beta_1\,\beta_0\,c_4[1,1]
     + \beta_2\, c_4[0,0,1]
     + \beta_0^2\,c_4[2]
     + \beta_1  c_4[0,1]
     + \beta_0\,c_4[1] \nonumber \\
   && \phantom{\beta_0^3\, c_4[3]+ \beta_1\,\beta_0\,c_4[1,1]+ \beta_0^2\,}~+c_4[0]\,,
       \label{eq:c_4} \\
   &\vdots& \nonumber \\
 c_{n}
   &=&\!\!\beta_0^{n-1}\, c_{n}[n\!-\!1]+ \cdots + c_n[0]\,,
\label{eq:c_n}
\end{eqnarray}
\end{subequations}
where  $\beta_i$ are the expansion coefficients of the QCD $\beta$-function presented in Appendix \ref{App:B},
\begin{equation}
\label{eq:beta}
\mu^2\frac{d a_s(\mu^2)}{d \mu^2}=
\beta(a_s)=-a_s^{2}(\mu^2) \sum_{i\geq 1} \beta_{i-1} a_s^{i-1}(\mu^2)\,.
\end{equation}
The decomposition in Eqs.(\ref{eq:d_beta}) contains complete knowledge about
$\alpha_s$-renormalization in each order of expansion for the RGI quantity $C_\text{Bjp}$.
It makes it possible to work on optimization of the perturbation series.
According to Eq.~(\ref{eq:d_beta}), the explicit form of the $\{\beta\}$-expansion for $C_\text{Bjp}$ within
the sequential BLM approach \cite{Kataev:2014jba} is
\begin{subequations}
 \label{eq:C1-3mod}
 \begin{eqnarray}
C_\text{Bjp}(a_s)
 =1&&+  a_s(-3C_F)  \\
&&+a_s^2(-3C_F)\cdot
          \left\{\frac{1}{3}{\rm C_A}-\frac{7}{2}{\rm C_F}+
          2 \cdot\beta_0 \right\}\label{eq:C2}\\
&&+a_s^3(-3C_F)\cdot
          \left\{\frac{115}{18}\cdot \beta_0^2 +
          \left(\frac{59}{12}-4\zeta_3\right)\cdot \beta_1
          \right. \nonumber \\
&&        \left.  -\left[\left(
\frac{215}{36}- 32 \zeta_3+\frac{40}{3}\zeta_5\right){\rm C_A}
    +\bigg(\frac{166}{9}- \frac{16}3\zeta_3\bigg){\rm C_F}\right]\cdot \beta_0
    \right. \nonumber \\
&&     \left. +
 \bigg( \frac{523}{36} - 72\zeta_3\bigg){\rm C_A^2}+\frac{65}{3}{\rm C_F C_A}+ \frac{\rm C_F^2}{2}
 \right\} \\
 && +a_s^4(-3C_F)\cdot c_4\, .
\end{eqnarray}
  \end{subequations}
The last known coefficient $c_4$ has the explicit form \cite{Baikov:2010je}
\begin{eqnarray}
\!\!\!\!\!\!\!\!\!\!c_1\cdot c_4&=&\! (-3C_F)\cdot \bigg\{C_A^3 \left(-\frac{4276}{27} \zeta_3+\frac{968}{9}\zeta_3^2-\frac{25090 }{27}\zeta_5-\frac{1540 }{3}\zeta_7+\frac{8004277}{2916}\right) \nonumber \\
&&\!+n_f T_r \left[C_A^2 \left(-\frac{236}{3}\zeta_3 -\frac{704}{9} \zeta_3^2+\frac{14840}{27} \zeta_5+\frac{560}{3} \zeta_7-\frac{1238827}{486}\right) \right.
\nonumber\\
&&\!\left. +C_A C_\text{F} \left(\frac{20624}{27}\zeta_3-\frac{4400}{27} \zeta_5-\frac{2240
}{3}\zeta_7+\frac{87403}{162}\right)+C_\text{F}^2 \left(-\frac{3608}{9} \zeta_3+\frac{4640}{9} \zeta_5-\frac{839}{27}\right) \right] \nonumber \\
&&\!  +C_A^2 C_F \left(-\frac{25456}{27} \zeta_3+\frac{22000
}{27}\zeta_5+\frac{6160}{3} \zeta_7-\frac{1071641}{648}\right)+C_A C_\text{F}^2 \left(\frac{7768 }{9}\zeta_3-\frac{16720}{9} \zeta_5+\frac{3707}{54}\right) \nonumber
\end{eqnarray}
\begin{eqnarray}
&&\! +
(n_f T_r)^2
   \left[C_A \left(\frac{688}{27} \zeta_3+\frac{128}{9} \zeta_3^2-\frac{320 }{9}\zeta_5+\frac{165283}{243}\right)+C_\text{F} \left(\frac{1060}{27}-\frac{928}{9} \zeta_3\right)\right]  +C_F^3
   \left(32 \zeta_3+\frac{4823}{24}\right) \nonumber \\
   &&\!-n_f\frac{16\dRR}{3 C_\text{F} d_R}(13+ 16 \zeta_3-40 \zeta_5)+\frac{16 \dRA}{3 C_\text{F}
   d_R}(3 - 4 \zeta_3 - 20 \zeta_5)-\frac{38720}{729}(n_f T_r)^3
   \bigg\}\,,
\end{eqnarray}
with the $SU_{c}(N)$-group fundamental fermion invariants
\begin{eqnarray}
\label{eq:inv}
&&T_r = \frac{1}{2}\, ; \; C_\text{F}= \frac{N^2-1}{2N}\, ; \; C_A =N\, ;\;
N_A = 2C_\text{F} C_\text{A} \equiv N^2-1\, ; \nonumber \\
&&d^{abc}d^{abc}= \frac{(N^2-4)N_A}{N}\, ; ~\dRA = \frac{N(N^2+6)}{48}N_A\, ; \nonumber \\
&&\dRR = \frac{N^4-6N^2+18}{96N^2}N_A\, ; ~ \dAA = \frac{N^2(N^2+36)}{24}N_A\, ,
\end{eqnarray}
where $d_R$ is the dimension of the quark color representation, $d_R = 3$ in QCD, and $n_f$ denotes the number of active flavors.
The explicit form of the $\beta$-expansion for $c_4$ is not known yet.
The numerical form of $C_\text{Bjp}(a_s)$ \cite{Baikov:2010je} reads
\begin{eqnarray}
 \label{eq:C1-4numer}
C_\text{Bjp}(a_s)&=&1-4\Big[ a_s + a_s^2\left(\frac{55}3-\frac{4}3n_f\right)+a_s^3\left(663.04-121.72n_f+2.84n_f^2\right)+ \nonumber \\
&& a_s^4\left(30684.6 - 7897.05n_f + 482.64n_f^2 - 6.64n_f^3\right)\Big]\,.
\end{eqnarray}

\section{ RG solutions for QCD charge}
 \label{App:B}
\textbf{1.}
Asymptotic freedom is  the basic feature of QCD as the theory of strong interactions \cite{Gross:1973id,Politzer:1973fx}.
This leading order prediction was quickly complemented by the corresponding
2-loop \cite{Caswell:1974gg,Jones:1974mm} and 3-loop \cite{Tarasov:1980au,Larin:1993tp} results.
The 4-loop result was obtained 17 years later \cite{vanRitbergen:1997va} and
here we stay on this level of accuracy.
The explicit expressions for the first coefficients of $\beta$ function expansion are
\begin{eqnarray}
    \beta_0 &=& \frac{11}{3}\,C_\text{A} - \frac{4}{3}\,T_r n_f
    \,;\qquad
    \beta_1 = \frac{34}{3}\,C_{\text{A}}^{2}
        - \left(4C_\text{F}
        + \frac{20}{3}\,C_\text{A}\right)T_r n_f ;\nonumber \\
    \beta_2 &=&   \frac{2857}{54} C_A^3
 +2 C_F^2 T_r n_f - \frac{205}{9} C_F C_A T_r n_f
 - \frac{1415}{27} C_A^2 T_r n_f
 + \frac{44}{9} C_F (T_r n_f)^2 \nonumber \\
 &&
 + \frac{158}{27} C_A (T_r n_f)^2\, ; \; \nonumber \\
\beta_3 &=&
C_A^4 \left( \frac{150653}{486} - \frac{44}{9} \zeta_3 \right)+
C_A^3 T_R n_f \left( - \frac{39143}{81} + \frac{136}{3} \zeta_3 \right)
+ C_F^2 T_R^2 n_f^2 \left( \frac{1352}{27} - \frac{704}{9} \zeta_3
\right) \nonumber \\
 && +C_A C_F T_R^2 n_f^2 \left( \frac{17152}{243} + \frac{448}{9} \zeta_3 \right)
 + C_A C_F^2 T_R n_f \left( - \frac{4204}{27} + \frac{352}{9} \zeta_3 \right) + \frac{424}{243} C_A
 T_R^3 n_f^3 \nonumber \\
&&+ C_A^2 C_F T_R n_f \left( \frac{7073}{243} - \frac{656}{9} \zeta_3 \right)
+ C_A^2 T_R^2 n_f^2 \left( \frac{7930}{81} + \frac{224}{9} \zeta_3
\right) + \frac{1232}{243} C_F T_R^3 n_f^3 \nonumber \\
&&+ 46 C_F^3 T_R n_f + n_f \dRANA \left( \frac{512}{9} - \frac{1664}{3} \zeta_3 \right) + n_f^2
\dRRNA \left( - \frac{704}{9} + \frac{512}{3} \zeta_3
\right) \nonumber \\
&&+ \dAANA \left( - \frac{80}{9} + \frac{704}{3} \zeta_3 \right).
 \label{eq:beta0&1&2}
\end{eqnarray}
The corresponding 3- and 4-loop RG equations for the coupling
$A=\beta_0\,\alpha_s/(4\pi)$ read
\begin{eqnarray}
 \frac{d A_{(3)}}{dt}
  = - A_{(3)}^2\left[1 + b_1\,A_{(3)}+ b_2\,A_{(3)}^2\right]
\label{eq:beta.new}
\end{eqnarray}
and
\begin{eqnarray}
 \frac{d A_{(4)}}{dt}
  = - A_{(4)}^2\left[1 + b_1\,A_{(4)}+ b_2\,A_{(4)}^2+b_3\,A_{(4)}^3\right]
 \text{~~with~}b_i
 \equiv
 \frac{\beta_i}{\beta_0^{i+1}} \, .
\label{eq:beta3.new}
\end{eqnarray}
\textbf{2.} The solution of this RG equation at the 2-loop level ($b_2=b_3=0$)
assumes the form
\begin{eqnarray}
 \label{eq:App-RGExact}
 \frac{1}{A_{(2)}} +
 b_1
     \ln\left[\frac{A_{(2)}}{1+b_1 A_{(2)}}
     \right] = t\,.
\end{eqnarray}
The exact solution of Eq.~(\ref{eq:App-RGExact}) can be expressed in
terms of the Lambert function $W(z)$ \cite{Corless:1996zz} (see also
\cite{Magradze:1999um}), defined by
\begin{eqnarray}
z=W(z)\, e^{W(z)}\,.
\end{eqnarray}
This solution has the form
\begin{eqnarray}
 \label{eq:App-Exactsolution}
 A_{(2)}(t) =
 -\frac{1}{b_1} \frac{1}{1+W_{-1}(z(t))}\, ,
 \end{eqnarray}
where
$z(t) =
\left(1/b_1\right) \exp\left(-1+i\pi-t/b_1\right)
$
and the branches of the  multivalued function $W$ are denoted by
$W_{k}$, $k=0,\pm 1,\ldots $
The second-iteration solution of Eq.~(\ref{eq:App-RGExact}), which provides sufficient accuracy, is
\begin{eqnarray}
  \frac{1}{A_{(2)}(t)} \to  \frac{1}{A_{(2)}^\text{it-2} (t)}
=
  t + b_1\ln\left[t + b_1 + b_1\ln\left(t + b_1 \right)\right]\, .
\end{eqnarray}

\textbf{3.}  The approximate solution of the renormalization-group equation in the 4-loop
of QCD \cite{Tanabashi:2018}, where the $\beta$-function is given by
Eq.~(\ref{eq:beta3.new}), assumes the asymptotic expansion
\begin{eqnarray}
A_{(4)}(t) &\simeq& \frac{1}{t} \, \Big[ \,
1-\frac{b_1 l}{t} + \frac{1}{t^2}\Big( b_1^2 (l^2-l-1) + b_2 \Big) \Big. \nonumber \\
&&
+ \frac{1}{2t^3}\Big( b_1^3 (-2l^3+5l^2+4l-1) - 6b_1b_2l + b_3 \Big)
+ \frac{1}{6t^4} \Big( b_1^2b_2 (2l^2-l-1) \Big. \nonumber \\
&&
\Big.\Big. + b_1^4 (6l^4-26l^3-9l^2+24l+7) - b_1b_3 (12l+1) + 10 b_2^2 \Big) \, \Big] \, ,
\label{eq:beta.new.4L}
\end{eqnarray}
where $l=\ln (t)$.

\bibliography{refs}
\bibliographystyle{apsrev}

\end{document}